\documentclass[a4paper]{jpconf}
\usepackage{amsmath}
\usepackage{amssymb}
\usepackage[dvipdfm]{graphicx}
\usepackage{wrapfig}
\usepackage{bm}
\usepackage{float}
\usepackage{graphicx}
\usepackage{pstricks}
\usepackage{pst-eps}
\begin{document}
\title{Beyond the Tao-Thouless limit of the fractional quantum Hall
effect: spin chains and Fermi surface deformation}

\author{Masaaki Nakamura$^1$, Zheng-Yuan Wang$^1$ and Emil J. Bergholtz$^2$}

\address{
$^1$Department of Physics, Tokyo Institute of Technology,
Tokyo 152-8551, Japan,\\
$^2$Max-Planck-Institut f\"{u}r Physik komplexer Systeme,
N\"{o}thnitzer Stra{\ss}e 38, D-01187 Dresden, Germany}


\begin{abstract}
 We discuss the relationship between the fractional quantum Hall effect
 in the vicinity of the thin-torus, a.k.a. Tao-Thouless (TT), limit and
 quantum spin chains.  
%
%
 We argue that the energetics of fractional quantum Hall states in Jain
 sequence at filling fraction $\nu=p/(2p+1)$ (and $\nu=1-p/(2p+1)$) in
 the lowest Landau level is captured by $S=1$ spin chains with $p$ spins
 in the unit cell.
 These spin chains naturally arise at sub-leading order in
 $\e^{-2\pi^2/L_1^2}$ which serves as an expansion parameter away from
 the TT limit ($L_1\rightarrow 0$). We also corroborate earlier results
 on the smooth Fermi surface deformation of the gapless state at
 $\nu=1/2$, interpolating between a state described by a critical
 $S=1/2$ chain and the bulk.
\end{abstract}

\section{Introduction}
Ultra cold electrons in two dimensions in a strong perpendicular
magnetic field, the quantum Hall (QH) system \cite{qh}, exhibits a
fascinating phase diagram including phases with fractionalized
excitations and topological order
\cite{Laughlin,haldane83,halperin84,jain89,mr}. Ever since its discovery
three decades ago, the QH system has inspired a huge amount of
experimental and theoretical effort, not least due to its richness in
phenomenology and mathematical structure. New developments include the
observation of the fractional quantum Hall effect in graphene
\cite{fqhgraphene} and ideas of applications in the context of
topological quantum computing \cite{qcomp}.  Moreover, it has been
realized that the theoretical description of a system of rapidly
rotating bosons is formally very similar to that of an electron gas in a
magnetic field \cite{boseQH}.

A key property of fractional quantum Hall (FQH) states is their
topological order \cite{toporder}. One consequence thereof is that their
physical properties are insensitive to smooth deformations of the
manifold on which we choose to study them. This fact has been exploited
in a series of recent studies of the interacting many-body problem in
the limit geometry of a thin torus, referred to as the Tao-Thouless (TT)
limit \cite{Bergholtz-K2005-8,Seidel-F-L-L-M,ttnonab,Bergholtz-H-H-K}
(see also Refs. \cite{Tao-T,Anderson,Rezayi-H} for precursory studies
and Ref. \cite{related} for some recent related approaches).

In this proceeding we study the FQH system beyond the TT limit and argue
that some characteristics that are not readily understood in the TT
limit states are however manifest in the leading quantum fluctuations
away from this limit. While several of the ideas presented here have
been published earlier
\cite{Bergholtz-K2005-8,Nakamura-B-S,Bergholtz-N-S,Wikberg-B-K}, we
corroborate these results and also present a number of new results.

We study the structure of excitation spectra of two-dimensional
electrons in a magnetic field with Coulomb interactions as a function of
the torus circumference, $L_1$, and contrast the cases of even and odd
denominator filling fractions. For $\nu=1/2$ we interpret these results
in terms of a deforming Fermi sea \cite{Bergholtz-K2005-8}. Then, by
extending the results of Refs. \cite{Nakamura-B-S,Bergholtz-N-S} for the
$\nu=1/3$ FQH state, we further establish that FQH systems in a Jain
sequence\cite{jain89}, which has filling factor $\nu=p/(2p+1)$ in the
lowest Landau level can be described by $S=1$ spin chains with $p$ sites
unit cell.

The remainder of this proceeding is organized as follows. In Section
\ref{model} we review the lattice describing interacting particles in a
Landau level on the torus and present numerical spectra thereof as a
function of the circumference, $L_1$.  In
Section \ref{beyond}, we briefly motivate and restate known results
about the energy spectra in the TT limit and then introduce spin models
that provide an intuitive understanding of the numerical data away from
this limit. In particular, these considerations shed light on why odd
denominator fractions remain gapped while even denominator states tend
to be gapless.

\section{One-dimensional description and numerical simulations}\label{model}
We consider a model of $N$ interacting electrons on a torus pierced by $N_s$ flux quanta. In the Landau gauge, a
complete basis of $N_s$ degenerate single-particle states in the lowest Landau
level, labeled by
$k=0,\ldots, N_s-1$, can be chosen as
\begin{equation}
 \psi_k(x)=
  (\pi^{1/2} L_1)^{-\tfrac{1}{2}}\sum_{n=-\infty}^{\infty}
  \e^{i (k_1+n L_2) x_1}
  \e^{-\tfrac{1}{2}(x_2+k_1+n L_2)^2},\label{psik}
\end{equation}
where $L_i$ are the circumferences of the torus, $x_i$ the corresponding
coordinates, and $k_1=2\pi k/L_1$ the momentum along the $L_1$-cycle.
We have set the magnetic length $l_{\rm B}\equiv \sqrt{\hbar/eB}$ equal
to unity.

In this basis, any translation-invariant two-dimensional two-body
interaction Hamiltonian assumes the following
model on a one-dimensional discrete lattice,
\begin{equation}
\mathcal{H}=\sum_{k>|m|}\hat V_{km},\quad
\hat V_{km}\equiv V_{km}
\sum_{i}
 c_{i+m}^{\dag}
 c_{i+k}^{\dag}
 c_{i+m+k}^{\mathstrut}
 c_{i}^{\mathstrut},
 \label{1D_model}
\end{equation}
where the matrix-element $V_{km}$ specifies the amplitude for a process
where particles with separation $k+m$ hop $m$ steps to a separation
$k-m$ (note that $m$ can be negative).  The number of the lattice sites
is fixed by the area, $N_s=L_1 L_2/2\pi$.  At the
filling $\nu=p/q$ the Hamiltonian commutes with the center-of-mass
magnetic translations \cite{haldane85} $T_1$ and $T_2^q$ along the
cycles, which implies, in particular, that the total momentum $K_1$
along the $L_1$-cycle is conserved modulo $N_s$ in this gauge.  In this
system (\ref{1D_model}), two conservation numbers are given as
\begin{equation}
T_1:
 \e^{i 2\pi K_1}
 =\exp\biggl(i\frac{2\pi}{N_s}\sum_{j=1}^{N_s} j n_j\biggr), \quad
T_2^q:
 \e^{i 2q\pi K_2/N_s},
\end{equation}
where $n_j\equiv c_j^\dag c_j^{\mathstrut}$ and $K_2=0,1,\cdots,N_s/q$.

\begin{figure}
\includegraphics[width=8cm]{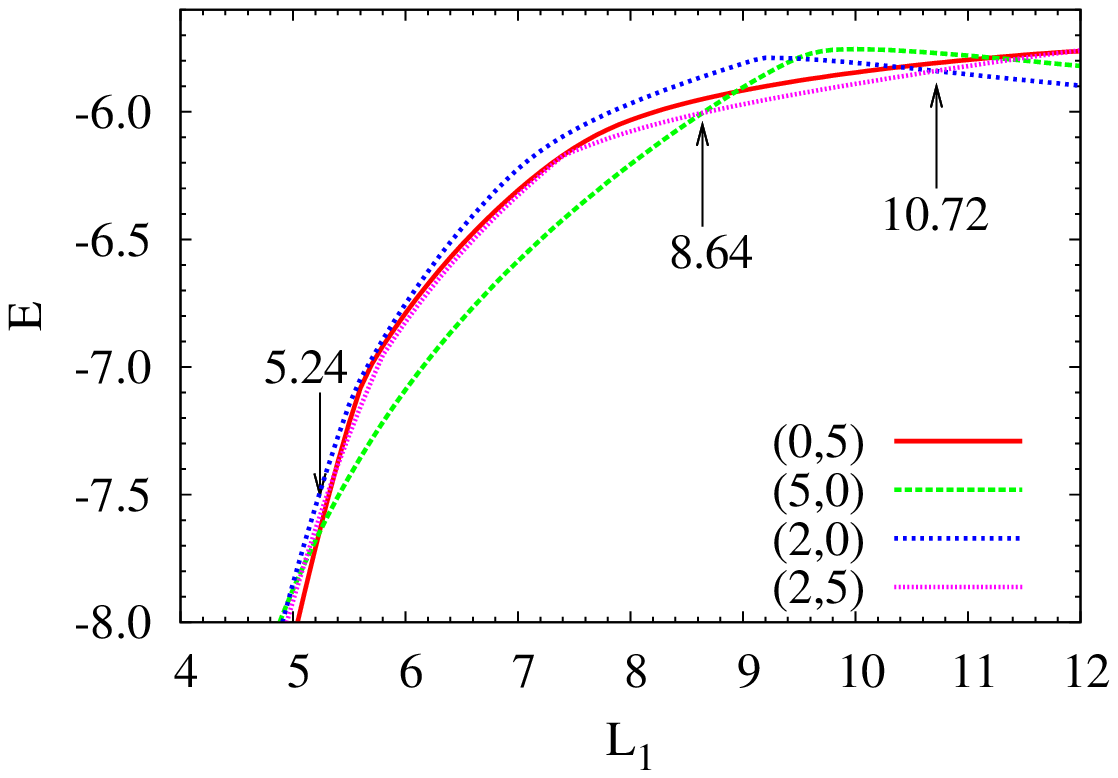}
\includegraphics[width=7cm]{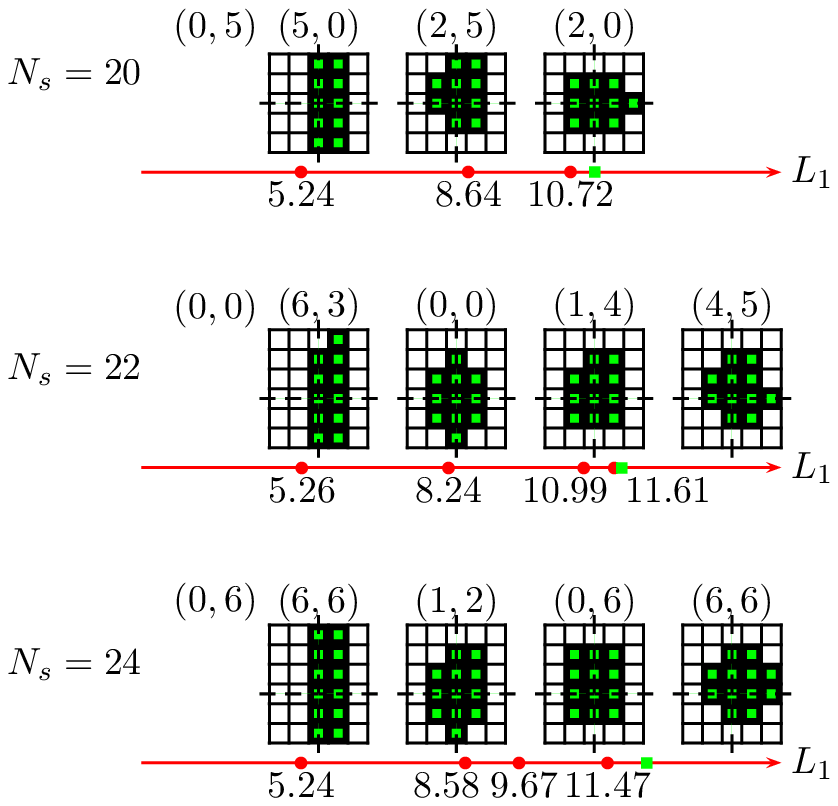}
\caption{The left panel shows energy spectra of the Hamiltonian
(\ref{1D_model}) with Coulomb potential for $\nu=1/2$ with $N_s=20$ as a
function of $L_1$. The corresponding phase diagram with ``Fermi sea''
representation of the gapless states labeled by quantum numbers
$(K_1,K_2)$ \cite{Bergholtz-K2005-8} is shown in the right panel along
with analogous phase diagrams for $N_s=22$ and $N_s=24$. The initial
phase transition at $L_1\approx 5.2-5.3$ as well as the level crossings
between the different Fermi seas are denoted by red circles while green
boxes denote the symmetric points $L_1=L_2$ which occur at $L_1=11.21$
($N_s=20$), $L_1=11.76$ ($N_s=22$) and $L_1=12.28$ ($N_s=24$).}
\label{spectra12}
\includegraphics[width=7.5cm]{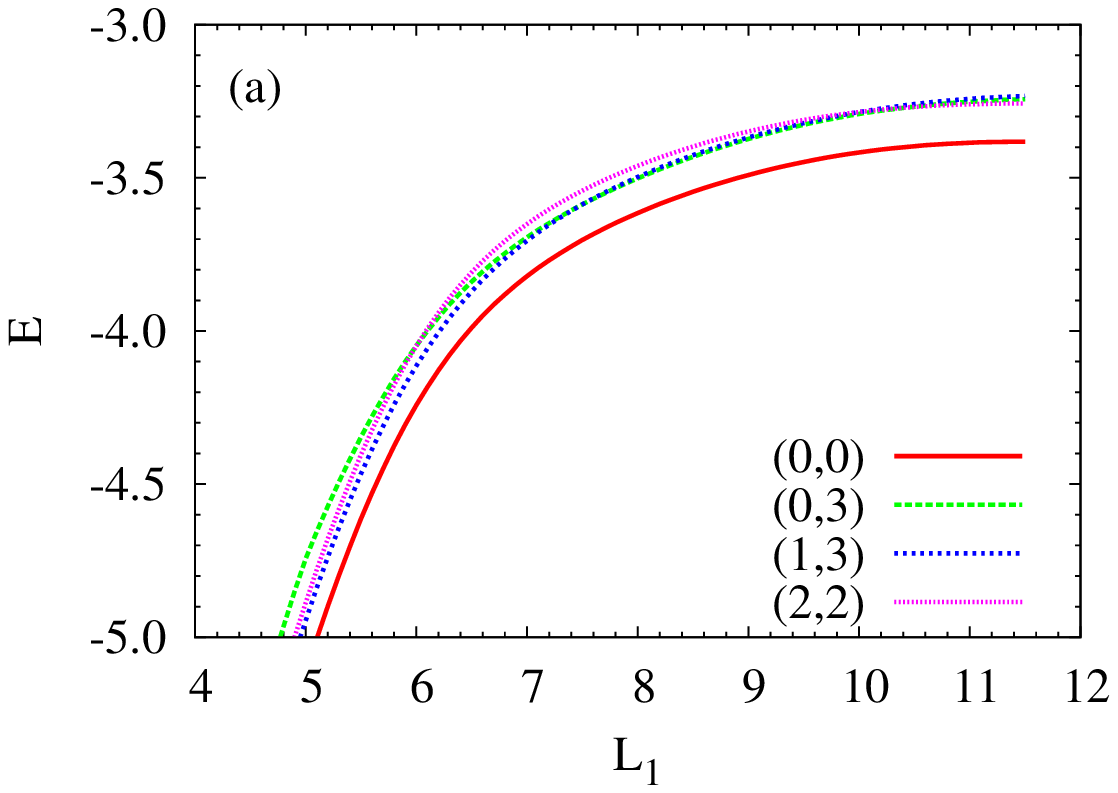}
\includegraphics[width=7.5cm]{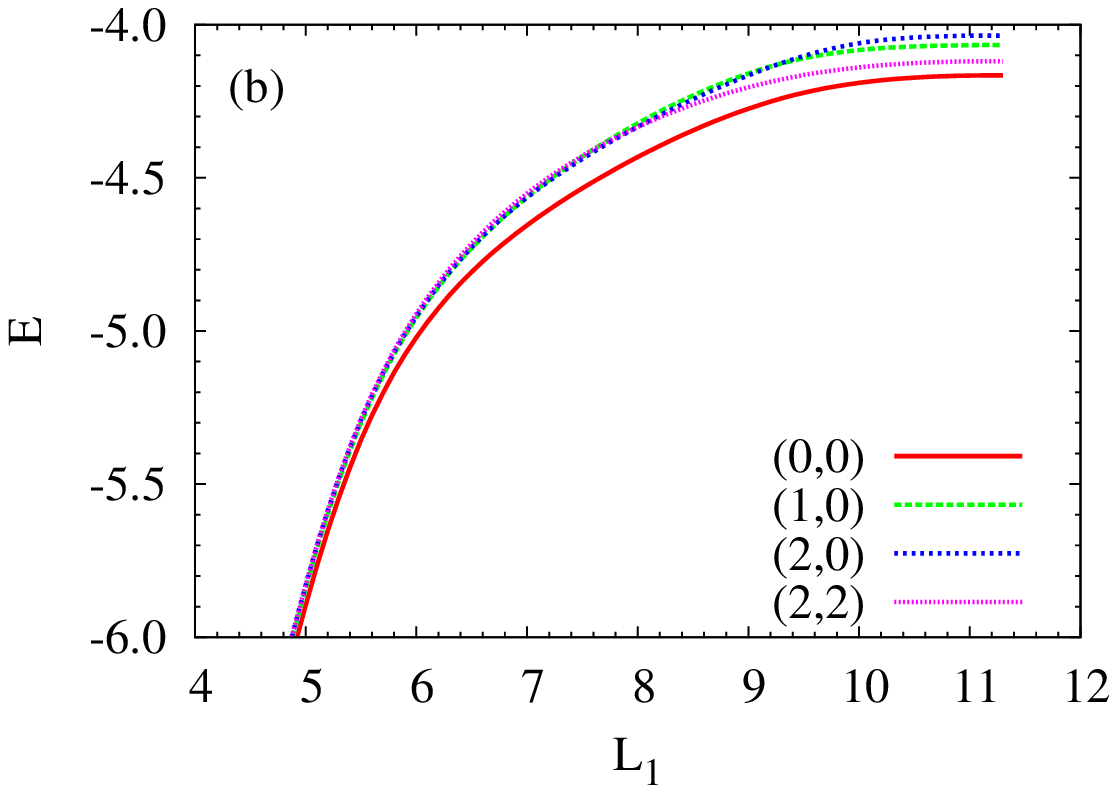}
\caption{Energy spectra of the Hamiltonian (\ref{1D_model}) for (a)
$\nu=1/3$ with $N_s=21$ and (b) $\nu=2/5$ with $N_s=20$. The energy gaps
never close as functions of the circumference $L_1$ of the torus.}
%
\label{spectra1325}
\end{figure}

We have calculated energy spectra of this system for $\nu=1/2,1/3,2/5$
using exact diagonalization in subspaces labeled by $(K_1,K_2)$. Using
these symmetries, the Hilbert space splits into $(N_s/q)^2$ sectors that
are all of comparable size. This enable us to treat larger systems.
The matrix elements of $V_{km}$ are calculated assuming Coulomb
interaction between the electrons. For the half-filled Landau level
$\nu=1/2$, we have calculated the energies up to $N_s=22$ which is
larger than the former analysis \cite{Bergholtz-K2005-8} (see
Fig.~\ref{spectra12}). Increasing $L_1$ from the TT limit ($L_1\to 0$),
the first level-crossing in the ground state energy takes place at
$L_1\simeq 5.3$ which corresponds to a phase transition from a
charge-density-wave (CDW) to a gapless phase. Further level crossings
occur among different gapless states, and finally $L_1$ arrives at the
duality point $L_1=L_2$ ($L_1=11.21$ for $N_s=20$, $L_1=11.76$ for
$N_s=22$ and $L_1=12.28$ for $N_s=24$). The spectrum is invariant under
$L_1\leftrightarrow L_2$ (although quantum numbers change) and hence we
only present data for $L_1\leq L_2$.
We have calculated the energies of the $\nu=1/3$
and $\nu=2/5$ cases, and confirmed that energy gaps do not close as
$L_1$ changes, and level-crossings occur only in excited states. This is
consistent with the fact that these states have energy gaps in the bulk
systems as is evident in the thin-torus limit (see below).

\section{Beyond the Tao-Thouless limit: Effective spin chains}\label{beyond}

\subsection{Tao-Thouless limit and leading quantum fluctuations}

From the shape of the single particle states (\ref{psik}) it follows that the distance between adjacent
states, the lattice constant $2\pi /L_1$, depends on
$L_1$ while the extent of the wave functions is constant. For small $L_1$ the overlap between different single
particle states decreases rapidly and the matrix elements $V_{km}$
simplify considerably. As $L_1\rightarrow 0$ one finds that 
\begin{equation}
V_{km}\sim \e^{-2\pi^2m^2/L_1^2}V_{k0},
\end{equation}
thus the $m\neq 0$ terms are exponentially suppressed for generic
interactions in this limit. The remaining ($m=0$) problem is exactly
solvable: ground states at any $\nu=p/q$ are gapped periodic crystals
(with a unit cell of $p$ electrons on $q$ sites) and the fractionally
charged excitations appear as domain walls between degenerate ground
states. The fractal structure (a.k.a. Devils staircase) of the Abelian
Haldane-Halperin hierarchy \cite{haldane83,halperin84} is manifest for
generic convex two-body interactions and provides an explicit estimate
for the relative stability of odd denominator states in excellent
agreement with experiment \cite{Bergholtz-H-H-K,hierarchyexp}.

The leading fluctuations are the $m=1$ terms describing correlated
hopping of two particles one site each in opposite directions. In a
series of earlier investigations, some of us have shown that these
fluctuations ultimately lead to a melting of the gapped state for
$\nu=1/2$ resulting in a gapless state \cite{Bergholtz-K2005-8}, while
for $\nu=1/3$ the excitation structure and gap is essentially unaltered,
even for very strong fluctuations
\cite{Nakamura-B-S,Bergholtz-N-S}. Below, we discuss how the numerical
results at finite $L_1$ can be interpreted in terms of effective spin
chains and extend this picture to include the most prominent ($m=1$)
Jain series, $\nu=p/(2p+1)$ and $\nu=(p+1)/(2(p+1)+1)$. The strategy is
as follows. First, we truncate the interaction to only include the
leading $m=1$ term, $\hat{V}_{21}$. This is especially sensible for a
pseudo potential interaction where
$V_{km}=P(k,m)\e^{-2\pi^2(k^2+m^2)/L_1^2}$ with $P(k,m)$ being a
polynomial\footnote{See Ref. \cite{Bergholtz-K2009} for explicit
expressions.} in $k$ and $m$. In this case our expansion in
$\e^{-2\pi^2/L_1^2}$ is well controlled, and we expect it to capture the
physics also for more general interactions. Second, we restrict the
local Hilbert space by including only local configurations that
correspond to the TT ground state and local applications of
$\hat{V}_{21}$. This naturally leads to a spin chain representation of
the Hamiltonian and enables us to deform it to connect to well studied
spin chain models and obtain physical insights by drawing analogies
between the physics of spin chains and FQH states.


\subsection{$S=1/2$ chain for $\nu=1/2$}
The electron system on a torus can be described in terms of $S=1/2$
variables in the following way \cite{Bergholtz-K2005-8}: In the vicinity
of the TT limit, the relevant physics is captured by a model containing
the three dominant terms as
\begin{equation}
  \mathcal{H}_t=\sum_{i}[V_{10}
  n_{i} n_{i+1}
  +V_{20}n_{i} n_{i+2}
  +V_{21}(c^{\dagger}_{i+1} c^{\mathstrut}_{i}
  c^{\dagger}_{i+2} c^{\mathstrut}_{i+3}
  +\mbox{H.c.})],
 \label{Truncated_Ham}
\end{equation}
where $V_{10}> V_{20}> V_{21}$.  In the TT
limit, the ground state is the charge-density-wave (CDW) state where
electrons are located on every two sites $|\cdots
010101010\cdots\rangle$, since the matrix elements of the electrostatic
term $V_{10}$ is dominant.  Away from the TT limit, the competition
between $V_{10}$, $V_{20}$ and $V_{21}$ can be included as an
interaction of the local spin states $|10\rangle\to|\uparrow\rangle$ and
$|01\rangle \to|\downarrow\rangle$ which means that
$c_{2n}^{\dagger}c_{2n+1}^{\mathstrut} \rightarrow S^+_n$,
$c_{2n+1}^{\dagger}c_{2n}^{\mathstrut} \rightarrow S^-_n$,
$c_{2n}^{\dagger}c_{2n}^{\mathstrut} \rightarrow 1/2+S^z_n$,
$c_{2n+1}^{\dagger}c_{2n+1}^{\mathstrut} \rightarrow 1/2-S^z_n$, where
$n$ is the index of the unit cell.  The effective spin Hamiltonian
is then given by the $S=1/2$ XXZ chain,
\begin{equation}
 \mathcal{H}_{XXZ}=\sum_{n}
  \frac{1}{2}(S_n^+S_{n+1}^-+S_n^-S_{n+1}^+)+\Delta S_n^zS_{n+1}^z,
  \label{XXZ_model}
\end{equation}
where $\Delta=(2V_{20}-V_{10})/4V_{21}$. It is well known that this
model has a phase transition from a ferromagnetic (CDW) phase to a
gapless (Tomonaga-Luttinger liquid) phase at $\Delta=-1$. This describes
the phase transition from the TT state to the gapless phase for the
half-filled Landau level $\nu=1/2$. The obtained solution corresponds to
the (first) elongated Fermi seas in Fig. \ref{spectra12} appearing for
$L_1\gtrsim5.3$.

\subsection{$S=1$ chain for $\nu=1/3$}
Next we consider spin-mapping for $\nu=1/3$ state.  In the TT limit, the
ground state of the system is the three-fold degenerate CDW state
$|\cdots\underline{010} \;
\underline{010}\cdots\rangle$ where the underlines denote unit cells.
To capture the effect of the leading quantum fluctuations, we introduce
$S=1$ spin variables as $|010\rangle \rightarrow
|0 \rangle$, $|100\rangle \rightarrow |+ \rangle$, $|001\rangle
\rightarrow |- \rangle$, where the operators are related as
\begin{equation}
 c_{3n}^{\dagger}c_{3n+1}^{\mathstrut}\to\frac{1}{\sqrt{2}}S^z_n S^+_n, \quad
 c_{3n+1}^{\dagger}c_{3n+2}^{\mathstrut}\to\frac{1}{\sqrt{2}}S^+_n S^z_n.
\label{1o3map}
\end{equation}
In this mapping, the three-fold degeneracy of the ground state is
hidden. The degenerate states have different center-of-mass quantum
numbers and there are no nonvanishing matrix elements between states in
different momentum sectors. It follows that there are three distinct
sectors which are described by the same effective Hamiltonian,
\begin{equation}
 \mathcal{H}_{1/3}=\sum_n 
  -\frac{V_{21}}{2}S^z_nS^+_nS^z_{n+1}S^-_{n+1} +\mbox{H.c.}
  =\sum_n
  \frac{V_{21}}{2}S^+_nS^-_{n+1}[1-(S^z_{n})^2][1-(S^z_{n+1})^2]+\mbox{H.c.}
  \label{S=1_chain_for_nu=1/3}
\end{equation}
This model is given by four spin interactions or the XY model with
projectors to the state $|\cdots 0000\cdots\rangle$.  Note that this
Hamiltonian does not have the space inversion and spin reversal
symmetries: the exchange process $|00\rangle\leftrightarrow|+-\rangle$
exists, but $|00\rangle\leftrightarrow|-+\rangle$ does not. The spin
rotational symmetry is also broken.  Actually, numerical analysis shows
that this Hamiltonian well describes the FQH state around $L_1 \simeq
7l$, by checking the overlap between the truncated Hamiltonian and full
Hamiltonian with Coulomb interaction and the Trugmann-Kivelson type
potential \cite{Trugman-K} which provides an exact parent Hamiltonian
for the Laughlin state. There is also a term given by a
function $\sum_n f(S^z_{n+1}-S^z_n)$ which stems from the electrostatic
terms, but this term does not alter any essential features of the present
model (\ref{S=1_chain_for_nu=1/3}) \cite{Wang-N-B}.  In
Refs.~\cite{Nakamura-B-S,Bergholtz-N-S}, this model with extra
parameters
\begin{align}
 \mathcal{H}(\lambda,D)
 &=\lambda\mathcal{H}_{1/3}
 +(1-\lambda)\mathcal{H}_{\rm Hei}(D),
 \label{S1extended}
 \\
 \mathcal{H}_{\rm Hei}(D)
 &=\sum_n \left\{ \bm{S}_n \cdot \bm{S}_{n+1}
 +D(S_n^z)^2  \right\}
\end{align}
has been analyzed numerically to study adiabatic continuity from a
conventional spin chain ($\lambda=0$).  For $\lambda=0$, it is known
that there is a phase transition at $D=D_c(\simeq 0.968)$ between the
Haldane ($D<D_c$) and large-$D$ ($D>D_c$) phases \cite{Chen-H-S}.  In
Refs. \cite{Nakamura-B-S,Bergholtz-N-S} it was shown that the ground
state of the model (\ref{S=1_chain_for_nu=1/3}) is smoothly connected
both to the Haldane phase and the large-$D$ phase without closing the
energy gap under the change of parameters. This indicates that the spin
model describing the $\nu=1/3$ FQH state has the nature of these two
phases simultaneously, and there are indeed some similarities between
the $\nu=1/3$ FQH effect and the $S=1$ Haldane-gap state as noticed
early on \cite{Girvin-A}. The reason for the absence of a phase
transition is due to the breaking of the discrete symmetries. According
to Ref.~\cite{Pollmann}, Haldane and large-$D$ phases are smoothly
connected if the dihedral group, time reversal and parity symmetries are
broken. Since the present model (\ref{S1extended}) with $\lambda\neq 0$
breaks these three symmetries, our result is consistent with this
general argument. Most properties of a finite chain of this model
(\ref{S1extended}) are similar to those of a usual large-$D$ phase where
no edge spins appear, however this is not inconsistent with the absence
of a phase transition \cite{Pollmann}.

\subsection{Jain fractions}
Let us now consider extentions of the above $S=1$ mapping to other
filling factors.  As discussed by Jain \cite{jain89}, $\nu=p/(2mp+1)$
FQH states may be described by the composite fermion picture, where $2m$
quantum flux are attached to non-interacting electrons of the $p$-th
Landau level, and projection onto the lowest Landau level. For
definiteness, we will consider the positive $m=1$ Jain sequence
$\nu=p/(2p+1)$, $p>0$. The results are easily transfered to the negative
Jain series by noting that our formulation is particle-hole symmetric,
hence the $\nu=p/(2p+1)$ is equivalent to the
$\nu=1-p/(2p+1)=(p+1)/(2(p+1)-1)$ case. This is in contrast to the wave
function based approaches and the composite fermion phenomenology, which
lack this symmetry. Before giving the general result, let us consider
the $\nu=2/5$ state ($p=2$) as the simplest example. In the TT limit,
the ground state is given by the CDW state with the configuration
$|\cdots \underline{01010}\;\underline{01010} \cdots \rangle$. When
$\hat{V}_{21}$ and $\hat{V}_{21}^{\dag}=\hat{V}_{2,-1}$ are applied to
this CDW state several times, one finds that the number of electrons in
each unit cell is always conserved, and the configuration $|\cdots 000
\cdots \rangle$ never appears. Therefore, configurations such as
$|\cdots \underline{00100} \cdots \rangle$, $|\cdots \underline{00011}
\cdots \rangle$, $|\cdots \underline{10001} \cdots \rangle$, $|\cdots
\underline{11000} \cdots \rangle$ are absent in the truncated Hilbert
space. This means that the $\nu=2/5$ states can be mapped to two $S=1$
variables by inserting $0$ appropriately (between the two $1$'s) in each
unit cell:
\begin{align}
 &
 |01010 \rangle \rightarrow | 01[00]10 \rangle \rightarrow |00
  \rangle,\\
 &
  | 00110 \rangle \rightarrow | 00[10]10 \rangle \rightarrow
  |\mathsf{-}0 \rangle,\\
 &
  | 01100 \rangle \rightarrow | 01[01]00 \rangle
  \rightarrow |0\mathsf{+} \rangle.
\end{align}
The relationship between fermion and spin-$1$ operators ($S_n^{\alpha}$,
$T_n^{\alpha}$) for the $\nu=2/5$ state is summarized in
Table~\ref{STmapping}.  We now obtain an effective spin chain as
\begin{align}
 \mathcal{H}_{2/5}
 =\frac{V_{21}}{2}\sum_n
 \left[
 T^{-}_n T_n^{z}S_n^{z}S^{+}_n+
 T_n^{z}T^{-}_nS^{+}_n S^{z}_n-
 S^{z}_nS^{-}_nT^{z}_{n+1}T^{+}_{n+1} \right]
 +\mbox{H.c.}
\end{align}

\begin{table}
\begin{center}
\begin{tabular}{|c|c||c|c|}
 \hline
 $c_{5n}^{\dagger}c_{5n+1}^{\mathstrut}$ &
 $2^{-1/2}T^{z}_nT^{+}_n$
 & $c_{5n+1}^{\dagger}c_{5n}^{\mathstrut}$ &
 $2^{-1/2}T^{-}_n T^{z}_n$\\
 $c_{5n+1}^{\dagger}c_{5n+2}^{\mathstrut}$ &
 $-2^{-1/2}T^{+}_n T^{z}_n$
 & $c_{5n+2}^{\dagger}c_{5n+1}^{\mathstrut}$
 & $ -2^{-3/2}T^{z}_nT^{-}_nS_n^-S_n^+$ \\
 $c_{5n+2}^{\dagger}c_{5n+3}^{\mathstrut}$
 & $2^{-3/2}S^{z}_n S^{+}_nT_n^+T_n^-$
 & $c_{5n+3}^{\dagger}c_{5n+2}^{\mathstrut}$
 & $2^{-1/2}S^{-}_n S^{z}_n$ \\
 $c_{5n+3}^{\dagger}c_{5n+4}^{\mathstrut}$
 & $-2^{-1/2}S^{+}_n S^{z}_n$
 & $c_{5n+4}^{\dagger}c_{5n+3}^{\mathstrut}$
 & $-2^{-1/2}S^{z}_n S^{-}_n$\\
 \hline
\end{tabular}
\end{center}
\caption{Relationship between fermion and spin-$1$ operators for the
$\nu=2/5$ state.}  \label{STmapping}
\end{table}

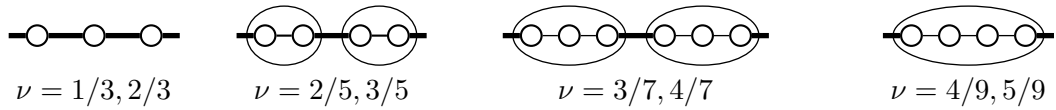
\begin{figure}[t]
\begin{center}
 \psset{unit=5mm,fillcolor=white}
  \begin{pspicture}(0,-2)(27,1)
   \rput(0,0){
   \rput[c](1.5,-1.5){$\nu=1/3, 2/3$}
   \pscircle(0,0){0.3}\pscircle(1.5,0){0.3}\pscircle(3,0){0.3}
   \psline[linestyle=solid,linewidth=2.0pt]{-}(0.3,0)(1.2,0)
   \psline[linestyle=solid,linewidth=2.0pt]{-}(1.8,0)(2.7,0)
   }
   \psline[linestyle=solid,linewidth=2.0pt]{-}(-0.75,0)(-0.3,0)
   \psline[linestyle=solid,linewidth=2.0pt]{-}(3.3,0)(3.75,0)
   \rput(6,0){
   \rput[c](1.75,-1.5){$\nu=2/5, 3/5$}
   \multirput(0,0)(2.5,0){2}{
   \psellipse[linewidth=0.5pt,hatchangle=0](0.5,0)(1.0,0.8)
   \pscircle(0,0){0.3}\pscircle(1,0){0.3}
   \psline[linestyle=solid]{-}(0.3,0)(0.7,0)
   }
   \psline[linestyle=solid,linewidth=2.0pt]{-}(-0.75,0)(-0.3,0)
   \psline[linestyle=solid,linewidth=2.0pt]{-}(1.3,0)(2.2,0)
   \psline[linestyle=solid,linewidth=2.0pt]{-}(3.8,0)(4.25,0)
   }
   \rput(13,0){
   \rput[c](2.75,-1.5){$\nu=3/7, 4/7$}
   \multirput(0,0)(3.5,0){2}{
   \psellipse[linewidth=0.5pt,hatchangle=0](1.0,0)(1.5,0.8)
   \pscircle(0,0){0.3}\pscircle(1,0){0.3}\pscircle(2,0){0.3}
   \psline[linestyle=solid,linewidth=0.5pt]{-}(0.3,0)(0.7,0)
   \psline[linestyle=solid,linewidth=0.5pt]{-}(1.3,0)(1.7,0)
   }
   \psline[linestyle=solid,linewidth=2.0pt]{-}(-0.75,0)(-0.3,0)
   \psline[linestyle=solid,linewidth=2.0pt]{-}(2.3,0)(3.2,0)
   \psline[linestyle=solid,linewidth=2.0pt]{-}(5.8,0)(6.25,0)
   }
   \rput(23,0){
   \rput[c](1.5,-1.5){$\nu=4/9, 5/9$}
   \multirput(0,0)(4,0){1}{
   \psellipse[linewidth=0.5pt,hatchangle=0](1.5,0)(2.0,0.8)
   \pscircle(0,0){0.3}\pscircle(1,0){0.3}
   \pscircle(2,0){0.3}\pscircle(3,0){0.3}
   \psline[linestyle=solid,linewidth=0.5pt]{-}(0.3,0)(0.7,0)
   \psline[linestyle=solid,linewidth=0.5pt]{-}(1.3,0)(1.7,0)
   \psline[linestyle=solid,linewidth=0.5pt]{-}(2.3,0)(2.7,0)
   }
   \psline[linestyle=solid,linewidth=2.0pt]{-}(-0.75,0)(-0.3,0)
   \psline[linestyle=solid,linewidth=2.0pt]{-}(3.3,0)(3.75,0)
   }
  \end{pspicture}
\end{center}
\caption{Effective $S=1$ spin chains for $\nu=p/(2p+1)$ ($p\geq 1$) FQH
 states given by $p$ spins in a unit cell. Mapping of negative series
 can also be possible via particle-hole transformation.}
 \label{S1_chain_p_unit}
\end{figure}

Similarly, we can extend the $S=1$ spin mapping for general $p$. For
$p>0$ cases, the configuration in the TT limit is given by $| \cdots
\underline{0(10)_{p}} \cdots \rangle$, and we obtain $S=1$ chains with
$p$-sites unit cell (see also Fig.~\ref{S1_chain_p_unit}),
\begin{equation}
\mathcal{H}_{\frac{p}{2p+1}}
=\frac{V_{21}}{2}\sum_n
\left[
\sum_{j=1}^{p-1}\left( 
S^{-}_{j,n} S_{j,n}^{z}S_{j+1,n}^{z}S^{+}_{j+1,n}+
S_{j,n}^{z}S^{-}_{j,n}S^{+}_{j+1,n} S^{z}_{j+1,n}
 \right)-
S^{z}_{p,n}S^{-}_{p,n}S^{z}_{1,n+1}S^{+}_{1,n+1}
\right]+\mbox{H.c.}
\label{S=1mapping}
\end{equation}
where $S_{j,n}^{\alpha}$ means $S=1$ operators of $j$-th site in $n$-th
unit cell. We have again ignored the contribution from
the electrostatic terms, for simplicity.  Note that this model includes the 
$\nu=1/3$ model in eq.~(\ref{S=1_chain_for_nu=1/3}), but is not appropriate for the $\nu=1/2$
state. Since this model also breaks the three discrete symmetries
mentioned above, we expect that it has similar low-energy properties as
$S=p$ quantum spin chains with Haldane gaps.  For $p\leq -2$ cases,
effective spin chains can be obtained by replacing $p$ in
eq.~(\ref{S=1mapping}) by $|p|-1$.  The detailed derivation and further
analysis of this model will be published elsewhere \cite{Wang-N-B}.

\section{Conclusion}
We have studied the fractional quantum Hall (FQH) effect based on the
one-dimensional description with torus boundary conditions and
illustrated how energy spectra of interacting electrons in a magnetic
field changes as functions of the circumference of the torus for various
filing factors using exact diagonalization. In an effort to understand
these results microscopically, we discussed spin chain mappings of the
FQH states. By extending earlier results for the $\nu=1/2$ and the
$\nu=1/3$ cases, we mapped the Jain sequences $\nu=p/(2p+1)$ and
$\nu=(p+1)/(2(p+1)-1)$ to $S=1$ chains with $p$ spins in the unit cell
that we conjecture to be gapped for all $p$. This further strengthens
the known analogies between the FQH physics and the quantum spin chains.
We have also studied the gapless state at $\nu=1/2$ numerically. By
going to slightly larger system sizes than before (from $10$ to $12$
particles), we confirmed that the evolution from the thin torus to the
bulk is, after an initial first order transition that is accurately
describe by an $S=1/2$ spin chain model, smooth and described by a
natural deformation of the quasi-particle Fermi sea.

\section{Acknowledgments}
We thank A.~Karlhede and J.~Suorsa for related collaborations, and
K.~Okamoto and M.~Oshikawa for discussions.  M. N. acknowledges support
from Global Center of Excellence Program ``Nanoscience and Quantum
Physics'' of the Tokyo Institute of Technology by MEXT.

\section{References}

\end{document}